\documentstyle[12pt,epsfig]{article}
\def\fo{\hbox{{1}\kern-.25em\hbox{l}}}

\newcommand{\newc}{\newcommand}

\newc{\lcal}{\int {\cal L}dt}

\newc{\LSP}{{\chi^0_1}}
\newc{\stauR}{{\tilde \tau_R}}
\newc{\stau}{{\tilde \tau_1}}
\newc{\mstop}{m_{\tilde{t}}}
\newc{\mHpm}{m_{H^\pm}}
\newc{\gsim}{\lower.7ex\hbox{$\;\stackrel{\textstyle>}{\sim}\;$}}
\newc{\lsim}{\lower.7ex\hbox{$\;\stackrel{\textstyle<}{\sim}\;$}}
\newc{\ie}{{\it i.e.}}
\newc{\etal}{{\it et al.}}
\newc{\eg}{{\it e.g.}}
\newc{\kev}{\hbox{\rm\,keV}}
\newc{\mev}{\hbox{\rm\,MeV}}
\newc{\gev}{\hbox{\rm\,GeV}}
\newc{\tev}{\hbox{\rm\,TeV}}
\newc{\xpb}{\hbox{\rm\, pb}}
\newc{\xfb}{\hbox{\rm\, fb}}

\newc{\mtop}{m_t}
\newc{\mbot}{m_b}
\newc{\mz}{m_Z}
\newc{\mw}{M_W}
\newc{\alphasmz}{\alpha_s(m_Z^2)}
\newc{\swsq}{\sin^2\theta_W}
\newc{\tw}{\tan\theta_W}
\newc{\cw}{\cos\theta_W}
\newc{\sw}{\sin\theta_W}
\newc{\BR}{\hbox{\rm BR}}
\newc{\zbb}{Z\to b\bar}
\newc{\Gb}{\Gamma (Z\to b\bar b)}
\newc{\Gh}{\Gamma (Z\to \hbox{\rm hadrons})}
\newc{\rbsm}{R_b^\hbox{\rm sm}}
\newc{\rbsusy}{R_b^\hbox{\rm susy}}
\newc{\drb}{\delta R_b}

\newc{\sgn}{\mbox{sgn}}

\newc{\tbeta}{\tan\beta}
\newc{\uL}{{\tilde u_L}}
\newc{\uR}{{\tilde u_R}}
\newc{\cL}{{\tilde c_L}}
\newc{\cR}{{\tilde c_R}}
\newc{\tL}{{\tilde t_L}}
\newc{\tR}{{\tilde t_R}}
\newc{\dL}{{\tilde d_L}}
\newc{\dR}{{\tilde d_R}}
\newc{\sL}{{\tilde s_L}}
\newc{\sR}{{\tilde s_R}}
\newc{\bL}{{\tilde b_L}}
\newc{\bR}{{\tilde b_R}}
\newc{\eL}{{\tilde e_L}}
\newc{\eR}{{\tilde e_R}}
\newc{\mhp}{m_{H^\pm}}
\newc{\mhalf}{m_{1/2}}
\newc{\emt}{{e/\mu /\tau}}

\newc{\lR}{\tilde{l}_R}
\newc{\lL}{\tilde{l}_L}
\newc{\nL}{\tilde{\nu}_L}
\newc{\na}{\chi^0_1}
\newc{\nb}{\chi^0_2}
\newc{\nc}{\chi^0_3}
\newc{\nd}{\chi^0_4}
\newc{\ca}{\chi^{\pm}_1}
\newc{\cb}{\chi^{\pm}_2}
\newc{\camp}{\chi^\mp_1}
\newc{\cbmp}{\chi^\mp_1}
\newc{\capos}{\chi^{+}_1}
\newc{\caneg}{\chi^{-}_1}
\newc{\phit}{\phi_t}
\newc{\phib}{\phi_b}
\newc{\phiew}{\phi_{ew}}
\newc{\htz}{h^0_t}
\newc{\hbz}{h^0_b}
\newc{\hewz}{h^0_{ew}}
\newc{\hsmz}{h^0_{sm}}
\newc{\huz}{h^0_u}
\newc{\hsusyz}{h^0_{susy}}

\def\mp{M_P}

\def\beq{\begin{equation}}
\def\eeq{\end{equation}}
\def\bea{\begin{eqnarray}}
\def\eea{\end{eqnarray}}
\def\slashchar#1{\setbox0=\hbox{$#1$}           % set a box for #1
   \dimen0=\wd0                                 % and get its size
   \setbox1=\hbox{/} \dimen1=\wd1               % get size of /
   \ifdim\dimen0>\dimen1                        % #1 is bigger
      \rlap{\hbox to \dimen0{\hfil/\hfil}}      % so center / in box
      #1                                        % and print #1
   \else                                        % / is bigger
      \rlap{\hbox to \dimen1{\hfil$#1$\hfil}}   % so center #1
      /                                         % and print /
   \fi}                                         %
\catcode`@=11
\long\def\@caption#1[#2]#3{\par\addcontentsline{\csname
  ext@#1\endcsname}{#1}{\protect\numberline{\csname
  the#1\endcsname}{\ignorespaces #2}}\begingroup
    \small
    \@parboxrestore
    \@makecaption{\csname fnum@#1\endcsname}{\ignorespaces #3}\par
  \endgroup}
\catcode`@=12

\begin{document}

\baselineskip=18pt

\begin{titlepage}
\begin{flushright}
IZTECH-P/2006-05
\end{flushright}

\begin{center}
\vspace{1cm}

{\Large \bf Non-Gravitating Scalars and Spacetime
Compactification}

\vspace{0.5cm}

{\bf Durmu{\c s} A. Demir and Beyhan Puli{\c c}e}

\vspace{.8cm}

{\it Department of Physics, Izmir Institute of Technology, IZTECH,
TR35430, Izmir, Turkey}

\end{center}
\vspace{1cm}

\begin{abstract}
\medskip
We discuss role of partially gravitating scalar fields, scalar
fields whose energy-momentum tensors vanish for a subset of
dimensions, in dynamical compactification of a given set of
dimensions. We show that the resulting spacetime exhibits a
factorizable geometry consisting of usual four-dimensional
spacetime with full Poincare invariance times a manifold of extra
dimensions whose size and shape are determined by the scalar field
dynamics. Depending on the strength of its coupling to the
curvature scalar, the vacuum expectation value (VEV) of the scalar
field may or may not vanish. When its VEV is zero the higher
dimensional spacetime is completely flat and there is no
compactification effect at all. On the other hand, when its VEV is
nonzero the extra dimensions get spontaneously compactified. The
compactification process is such that a bulk cosmological constant
is utilized for curving the extra dimensions.

\end{abstract}

\bigskip
\bigskip

\begin{flushleft}
IZTECH-P/2006-05 \\
February 2006
\end{flushleft}

\end{titlepage}

%%%%%%%%%%%%%%%%%%%%%%%%%%%%%%%%%%%%%%%%%%%%%%%%%%%%%%%%%%%%%%%
%\tableofcontents

\section{Introduction}

The pioneering work of Kaluza \cite{kaluza}, which states that Einstein's gravity and electromagnetism can be
unified into five-dimensional general relativity if the extra dimension is barred from affecting laws of physics,
has given rise to, via Klein's compactification approach \cite{klein,einstein}, a wealth of higher-dimensional
theories motivated by various physical phenomena (see the review \cite{review}). For instance, a unified
quantum-theoretic description of gravity and other forces of Nature ($i.e.$ supergravity and superstring
theories) cannot be formulated without introducing extra dimensions (see $e.g.$ \cite{review2}). Moreover, extra
dimensions can provide a viable solution to gauge hierarchy problem of the standard electroweak theory with no
need to low-energy supersymmetry \cite{nima}. In these and many other applications the extra dimensions are
assumed to roll up to form a sufficiently small space. The characteristic size of the extra dimensions can vary
from Planck length up to a few $mm$, as allowed by the present experimental bounds.

The aforementioned scheme is incomplete in that it is necessary to explain how or why extra dimensions differed
so markedly in size and topology from the ordinary four dimensions. In other words, it is necessary to find the
dynamical mechanism that leads to compactification of the extra dimensions, that is, the macroscopic
four-dimensional spacetime times the compact manifold of extra dimensions should be an energetically-preferred
solution of the higher-dimensional Einstein equations. This has been accomplished by utilizing higher-curvature
gravity \cite{modify} and by coupling Einstein gravity to matter in a judicious way. The latter leads to
spontaneous compactification of extra dimensions as was first pointed out in \cite{first}. Spontaneous
compactification has been realized with Yang-Mills fields \cite{nonabel}, antisymmetric tensor fields \cite{sc},
sigma model fields \cite{gz,gz1}, and conformally-coupled scalars \cite{nilles}. In each case, components of the
Ricci tensor are balanced by those of the stress tensor, and depending on the structure of the latter a subset of
dimensions are compactified.

In this work we are interested in dynamical compactification induced by scalar fields. The role of scalars in
dynamical compactification process was first analyzed in \cite{gz,gz1} where a $D$--dimensional minimally-coupled
non-linear sigma model with metric $h_{i j}(\phi)$ ($i,j=1,\dots,D$) was shown to lead to a dynamical
compactification of $D$ extra dimensions provided that sigma model metric is Einstein. In other words, equations
of motion for the metric field requires the Ricci tensor ${\cal{R}}_{i j}$ to be proportional to $h_{i j}(\phi)$,
and thus, $D$-dimensional extra space relaxes to the geometry of the sigma model. The remaining dimensions
$x^{\mu}$ ($\mu=0,1,\dots$) span a strictly flat Minkowski space. That this set-up compactifies the extra
dimensions $y^{i}$ becomes especially clear with the ansatz $\phi^i = y^i$ or any function of $y^i$. In the
literature role of scalars in spontaneous compactification was also emphasized in \cite{nilles} where a
strictly-conformal invariant scalar-tensor theory (though conformal invariant Weyl contribution is absent) of
gravity in six dimensions is shown to lead to compactification of the two extra dimensions and simultaneously
generate Newton's constant spontaneously. The scalar fields can be chosen in varying group representations
depending on the desired compactification structure, and the gauge symmetry in four dimensions turns out to be
smaller than the isometries of the compact manifold.

Obviously, the space of extra dimensions may \cite{first,nonabel,gz,gz1} or may not \cite{gz1,noncompact} form a
compact space. The extra dimensions can possess negative curvature \cite{arefeva} yet they can still be compact
\cite{bize}. Moreover, the geometry does not need to be factorizable \cite{rs}. In general, shape and topology of
the extra space are entirely determined by the mechanism of dynamical compactification.

In this work we discuss yet another compactification mechanism induced by scalar fields. We will show that a
single scalar field living in a higher dimensional spacetime can lead to dynamical compactification of the extra
dimensions without inducing a classical cosmological constant when it gravitates only in those dimensions which
are to be compactified. For proving this statement, it is necessary to show first that a strictly flat spacetime
supports non-trivial scalar field configurations. This we will do in Sec. II. The next step is to show the
compactification of the extra dimensions into a $D$--dimensional manifold, and this we will show in Sec. III. We
will conclude in Sec. IV.

\section{Partially Gravitating Scalar Fields}
Let us consider a real scalar field $\phi$ living in a $(4+D)$--dimensional spacetime with coordinates
$z^{A}=\left(x^{\mu}, y^i\right)$ where $\mu=0,\dots, 3$ and $i=1,\dots, D$. Keeping the gravitational sector
minimal, the most general action integral takes the form
\begin{eqnarray}
\label{newaction} S&=&\int d^{4+D}z\, \sqrt{-g} \Bigg\{ \frac{1}{2} M_{\star}^{D+2} {\cal{R}}\nonumber\\ &-&
\frac{1}{2} g^{A B} \partial_A \phi\,
\partial_B \phi - \frac{1}{2} \zeta {\cal {R}} \phi^2 - V(\phi)\Bigg\}
\end{eqnarray}
where we have adopted $\left(-1, +1, \dots, +1\right)$ metric signature, and denoted the curvature scalar by
${\cal{R}}$ and fundamental scale of gravity by  $M_{\star}$. There is no symmetry principle\footnote{The
Goldstone bosons generated by spontaneously broken continuous symmetries is an exception. They do not couple to
the curvature scalar directly \cite{voloshin,demir}.} for avoiding direct coupling of $\phi$ to the curvature
scalar, namely, a scalar field should always exhibit $\zeta {\cal {R}} \phi^2$ type interaction with Ricci
scalar. Note that the scalar field theory in (\ref{newaction}) exhibits conformal invariance when $V(\phi)
\propto \phi^{4+D}$ and $\zeta = \zeta_{4+D}$, where
\begin{eqnarray}
\zeta_{4+D} = \frac{D+2}{4 \left(D+3\right)}
\end{eqnarray}
which equals $1/6$ for $D=0$ and $1/4$ for $D=\infty$. This property, however, is of little use since conformal
invariance is explicitly broken by the Einstein-Hilbert term, and full invariance cannot be achieved unless it is
replaced by a term quadratic in Weyl tensor \cite{demir,ci}.

The action (\ref{newaction}) attains its extremum when scalar and metric fields obey their equations of motion
\begin{eqnarray}
{\cal{R}}_{A B} &=& \frac{{\cal{T}}_{A B}(\phi)}{M_{\star}^{D+2}-\zeta \phi^2} \label{einstein}\\
\Box \phi &=& \zeta R \phi + V^{\prime}(\phi) \label{phieqn}
\end{eqnarray}
where prime denotes differentiation with respect to $\phi$. Here, the source term for the Ricci tensor is given
by
\begin{eqnarray}
\label{yenitab}
{\cal{T}}_{A B}(\phi) &=& T_{A B} - \frac{1}{D+2}  g_{A B} g^{C D} T_{C D}\nonumber\\
                      &=& \partial_{A}\phi \partial_B \phi - \zeta \nabla_A \nabla_B  \phi^2 \nonumber\\
                      &+& \frac{1}{D+2} \left( 2 V(\phi) - \zeta \Box \phi^2\right) g_{A B}
\end{eqnarray}
where
\begin{eqnarray}
\label{tab} T_{A B} &=& \partial_{A}\phi \partial_B \phi  - g_{A B} \left( \frac{1}{2} g^{C D} \partial_C \phi
\partial_D \phi + V(\phi)\right)\nonumber\\&+& \zeta \left( g_{A B} \Box - \nabla_A \nabla_B \right) \phi^2
\end{eqnarray}
is the stress tensor of $\phi(z^A)$. The second line at right-hand side follows from direct coupling of $\phi$ to
Ricci scalar, and it remains non-vanishing even in the flat limit \cite{energymomentum}. One can show that
$\nabla^A T_{A B} = - \zeta {\cal{G}}_{C B} \nabla^C \phi^2$ where
\begin{eqnarray}
{\cal{G}}_{A B} = {\cal{R}}_{A B} - \frac{1}{2} {\cal{R}} g_{A B}
\end{eqnarray}
is the Einstein tensor.  It is clear that $T_{A B}+\zeta \phi^2 {\cal{G}}_{A B}$ is a conserved tensor source in
agreement with the Bianchi identity.

The Einstein equations for the Ricci tensor, equation (\ref{einstein}), guarantee that if ${\cal{T}}_{A B}$
vanishes for a certain range of its indices so does ${\cal{R}}_{A B}$ for the same index ranges. When
${\cal{R}}_{A B}$ vanishes for a range of indices the metric tensor on that block will be assumed to be $\eta_{A
B}$. Therefore, a given range of indices for which ${\cal{R}}_{A B}=0$ will be interpreted to form a flat
subspace in a $(4+D)$--dimensional spacetime. In the opposite case, when ${\cal{T}}_{A B}$ is nonvanishing for a
range of indices so is ${\cal{R}}_{A B}$, and precise forms of the metric and scalar field are determined from a
self-consistent solution of (\ref{einstein}) and (\ref{phieqn}). In the following we discuss under what
conditions ${\cal{T}}_{A B}$ possesses specific texture zeroes.

\subsection{Non-gravitating Scalar Field}
We start our analysis by considering first a completely non-gravitating scalar $i.e.$ we impose ${\cal{T}}_{A B}
= 0 $ for all $A=(\mu, i)$ and $B=(\nu, j)$. This implies that ${\cal{R}}_{A B}$ vanishes for all $A,B$ so that
metric tensor may be assumed to take the form $\eta_{A B} = \left(-1, 1, \dots, 1\right)$, as mentioned
before\footnote{The purely non-gravitating scalar field configuration discussed in this subsection is not new at
all; it has been shown to exist already in \cite{nongrav} where one can find a more detailed description of the
solution of $T_{A B} =0$. This subsection is included here for completeness of the discussions.}. It is
convenient to nullify first ${\cal{T}}_{A B}$ for $A\neq B$. These equations receive contributions from the first
line of the second equality in (\ref{yenitab}) only, and they enforce $\phi$ to have the form
\begin{eqnarray}
\phi = \psi^{- \frac{2 \zeta}{1 - 4 \zeta}}
\end{eqnarray}
where  $\psi$ is another real scalar field. Then vanishing of the
diagonal entries of ${\cal{T}}_{A B}$ further determines $\psi$ to
be a second order polynomial in $z^{A}$, and $V(\phi)$ to be a
function of $\phi$ only. Consequently, one finds for $\phi(z)$
\begin{eqnarray}
\label{phisoln} \phi(z)\equiv \phi_{0}(z)=\left(
\frac{\tilde{a}}{2} \eta^{A B} z_{A} z_{B} +  \eta^{A B} z_{A}
\tilde{p}_{B} + \tilde{b}\right)^{-\frac{ 2 \zeta}{1- 4 \zeta}}
\end{eqnarray}
where $\tilde{a}$, $\tilde{b}$ and $\tilde{p}_{A}$ are constants
of integration. For $\phi(z)$ to take this rather specific form
its self-interaction potential must equal
\begin{eqnarray}
\label{pot} V(\phi_0) &=& 16 \tilde{a} (D+3)
\frac{\zeta^2}{\left(1- 4 \zeta\right)^2} \left(\zeta -
\zeta_{4+D}\right) \phi_0^{\frac{1}{2 \zeta}}
   \nonumber\\&+& 2 \left( \eta^{A B} \tilde{p}_{A} \tilde{p}_{B} - 2 \tilde{a} \tilde{b}\right)   \frac{\zeta^2}{\left(1- 4 \zeta\right)^2}
   \phi_0^{\frac{1 -2 \zeta}{\zeta}}
\end{eqnarray}
which explicitly depends on the parameters of (\ref{phisoln}).
Consequently, for ${\cal{T}}_{A B}$ to vanish the scalar field
itself does not need to vanish; all that is required is to devise
a self-interaction potential (\ref{pot}) on the specific solution
(\ref{phisoln}) for $\phi(z)$. One notices that this
non-gravitating nontrivial field configuration arises thanks to
the $\zeta$ dependent terms in ${\cal{T}}_{A B}$ or equivalently
the non-minimal coupling of $\phi$ to the curvature scalar.
Indeed, when $\zeta \rightarrow 0$ the scalar field reduces to a
constant and $V(\phi) \rightarrow 0$, which is a trivial
configuration.

It is not hard to see that (\ref{phisoln}) and (\ref{pot}) also nullify $T_{A B}$, the true energy-momentum
tensor of $\phi$ in (\ref{tab}). Actually, this coincidence is expected since the Einstein tensor vanishes
whenever the Ricci tensor vanishes. The fact that a non-minimally coupled scalar field possesses a non-trivial
configuration despite its vanishing $T_{AB}$ has recently been discussed in \cite{nongrav}, and field and
potential solutions in (\ref{phisoln}) and (\ref{pot}) have already been obtained therein. The solution for
$\phi$ in (\ref{phisoln}) represents a shock wave propagation. The wave front is spherical for $\tilde{p}_A = 0$
and planar for $\widetilde{a}=0$. When $\zeta = \zeta_{4+D}$ the first term in potential drops out, and the
second term becomes proportional to ${\phi}_0^{-(D+4)}$, which is precisely what is required by conformal
invariance \cite{demir,ci}.

An interesting property of the potential function (\ref{pot}) is
that its minimum varies with $\zeta$. Indeed, for $\zeta >
\zeta_{4+D}$ it is minimized at $\phi =0$ whereas its minimum
occurs at
\begin{eqnarray}
\label{mini} \overline{\phi}=\left( \frac{(D+3) (\zeta -
\zeta_{4+D})}{2 \zeta -1 } \frac{ 4 \tilde{a}}{\eta^{A B}
\tilde{p}_{A} \tilde{p}_{B} - 2 \tilde{a} \tilde{b}}\right)^{
\frac{2 \zeta}{1 - 4 \zeta}}
\end{eqnarray}
when $\zeta < \zeta_{4+D}$ and $\eta^{A B} \tilde{p}_{A}
\tilde{p}_{B} - 2 \tilde{a} \tilde{b} > 0$. In this sense the
conformal value of $\zeta$ represents a threshold point below and
above which the lowest energy configuration for $V(\phi_0)$
drastically changes.

So far we have discussed only the solution of ${\cal{T}}_{A B} = 0$ with no mention of the equation of motion of
$\phi$ in (\ref{phieqn}). Actually, the field configuration (\ref{phisoln}) with $V(\phi)$ given in (\ref{pot})
automatically satisfies (\ref{phieqn}). This observation is correct for all parameter ranges; in particular, at
the two possible minima of the potential: $\phi =0$ and $\phi = \overline{\phi}$. Therefore, it is not necessary
to require $\phi$ to take nonconstant values as claimed in \cite{nongrav}.

\subsection{Partially Gravitating Scalar Field}
In this section we discuss cases where $\phi$ gravitates only in a subset of dimensions. The construction of
completely non-gravitating scalar above will serve as a useful guide for our analysis. We will look for metric
and scalar field configurations in agreement with the following ${\cal{T}}_{A B}$ texture:
\begin{eqnarray}
{\cal{T}}_{\mu \nu}(\phi) &=& 0 \label{tmunu}\\ {\cal{T}}_{\mu j}(\phi) &=& {\cal{T}}_{i \nu}(\phi) = 0 \label{tmui}\\
{\cal{T}}_{i j}(\phi) &\neq& 0 \label{tij}
\end{eqnarray}
where ${\cal{T}}_{i j}(\phi)$ determines topology and shape of the extra space via (\ref{einstein}). As mentioned
before, when ${\cal{T}}_{A B}$ vanishes for a certain range of indices so does the Ricci tensor. This, however,
is not a trivial condition when $\phi$ gravitates in a subset of dimensions only. To clarify this point consider,
for instance, the constraint (\ref{tmunu}) above. It guarantees that ${\cal{R}}_{\mu\nu} = 0$; however, it cannot
guarantee, even with $g_{\mu \nu} = \eta_{\mu \nu}$, that the quartet $\left(x_0, x_1, x_2, x_3\right)$ forms a
flat space. The reason is that $\nabla_{\mu} \nabla_{\nu} \phi^2 = \partial_{\mu}\partial_{\nu} \phi^2$ if and
only if the connection coefficients, $\Gamma^{A}_{B C}$, satisfy $\Gamma^{A}_{\mu \nu} =0$ for all $(A,
\mu,\nu)$. This is guaranteed if $g_{\mu j}$ and $g_{i \nu}$ depend only on the extra dimensions. On the other
hand, considering $T_{\mu j}$ and $T_{i \nu}$, one finds that $\nabla_{\mu} \nabla_{i} = \partial_{\mu}
\partial_{i}$ if $g_{\mu j}$ and $g_{i \nu}$ both are constants with respect to all coordinates $x^{A}$, and if
$g_{i j}$ depends only on the extra dimensions. These flatness conditions on different groups of coordinates
implies that the metric tensor $g_{A B}$ must conform to structure of ${\cal{T}}_{A B}$ in
(\ref{tmunu}-\ref{tij}):
\begin{eqnarray}
g_{\mu \nu} &=& \eta_{\mu \nu} \label{gmunu}\\ g_{\mu j} &=& g_{i \nu} = 0 \label{gmui}\\ g_{i j} &=& g_{i
j}(\vec{y}) \label{gij}
\end{eqnarray}
which exhibits a block-diagonal structure as it should for extra coordinates $\left\{y^i\right\}$ to be
compactified $i.e.$ decoupled from the rest. With this structure for the metric tensor,  the source term of the
Ricci tensor ${\cal{R}}_{\mu\nu}$ in four dimensions takes the form
\begin{eqnarray}
\label{yenitmunu} {\cal{T}}_{\mu \nu}(\phi) &=& \partial_{\mu}\phi \partial_{\nu} \phi -
\zeta \partial_{\mu} \partial_{\nu}  \phi^2 \nonumber\\
                      &+& \frac{1}{D+2} \left( 2 V_{{\small \mbox{new}}}(\phi) -
                      \zeta \eta^{\alpha \beta} \partial_{\alpha} \partial_{\beta} \phi^2\right) \eta_{\mu \nu}
\end{eqnarray}
as follows from (\ref{yenitab}). Hence, as seen from four
dimensions, the self-interaction potential of $\phi$ is not the
original one $V(\phi)$, but
\begin{eqnarray}
\label{Vnew} V_{{\small \mbox{new}}}(\phi) &=& V(\phi) -
\frac{1}{2} \zeta g^{i j} \nabla_i \nabla_j \phi^2
\end{eqnarray}
which involves derivatives of $\phi^{2}$ with respect to extra
coordinates $\left\{y^i\right\}$. For ${\cal{T}}_{\mu \nu}(\phi)$
to vanish, first of all, the scalar field must have the special
form
\begin{eqnarray}
\label{phisoln2} \phi(z)\equiv \phi_0(z)=\left( \frac{a}{2}
\eta^{\mu \nu} x_{\mu} x_{\nu} +  \eta^{\mu \nu} x_{\mu} p_{\nu} +
b\right)^{-\frac{ 2 \zeta}{1- 4 \zeta}}
\end{eqnarray}
in analogy with (\ref{phisoln}) derived in Subsection A, above.
Here, in principle, all the parameters $a$, $b$ and $p_{\mu}$ are
functions of the extra coordinates $\left\{y^i\right\}$, and their
mass dimensions are $2 - (1-4 \zeta)(D+2)/4 \zeta$, $-(1- 4
\zeta)(D+2)/4 \zeta$ and $1 - (1-4 \zeta)(D+2)/4 \zeta$,
respectively. The scalar field configuration (\ref{phisoln2})
describes a shock wave propagation in four dimensions at each
point $\left\{y^i\right\}$ of the extra space.

Having $\phi(z)$ obeying to (\ref{phisoln2}) is not sufficient for
nullifying all components of ${\cal{T}}_{\mu \nu}$, however.
Indeed, for ${\cal{T}}_{\mu \nu}$ to vanish completely the
self-interaction potential felt by $\phi_0(z)$ must have the
special form
\begin{eqnarray}
\label{tildedV} \widetilde{V}(\phi_0) &=& 8 {a} (D+6)
\frac{\zeta^2}{\left(1- 4 \zeta\right)^2} \left(\zeta -
\zeta_{crit}\right) \phi_0^{\frac{1}{2 \zeta}}
   \nonumber\\&+& 2 \left( \eta^{\mu \nu} \tilde{p}_{\mu} \tilde{p}_{\nu} - 2 {a} {b}\right)   \frac{\zeta^2}{\left(1- 4 \zeta\right)^2}
   \phi_0^{\frac{1 -2 \zeta}{\zeta}}
\end{eqnarray}
which is to be contrasted with the potential function (\ref{pot})
of purely non-gravitating scalar field discussed in Sec. 2.1. The
most important difference between the two potentials comes from
replacement of $\zeta_{4+D}$ in (\ref{pot}) by
\begin{eqnarray}
\zeta_{crit} = \frac{(D+4)}{4(D+6)}
\end{eqnarray}
which ranges from $1/6$ at $D=0$ to $1/4$ at $D= \infty$. These
two critical $\zeta$ values, $\zeta_{crit}$ and $\zeta_{4+D}$,
agree at $D=0$ and $D=\infty$, but behave differently in between.
Clearly, $\zeta_{crit}$ arises from $1/(D+2)$ factor in
(\ref{yenitmunu}), and the two potentials (\ref{pot}) and
(\ref{tildedV}) coincide when $D=0$. In other words,
(\ref{yenitmunu}) is not the true stress tensor of a scalar field
living in four-dimensions; it is just cross section of the stress
tensor of a scalar field living in $(4+D)$ upon four-dimensional
subspace. It is with the special solution (\ref{tildedV}) $i.e.$
it is with
\begin{eqnarray}
\label{dotequal} V_{{\small \mbox{new}}}(\phi_0) =
\widetilde{V}(\phi_0)
\end{eqnarray}
which holds on $\phi(z)=\phi_0(z)$ that all ten components of
${\cal{T}}_{\mu \nu}$ and hence those of ${\cal{R}}_{\mu \nu}$
vanish with a strictly flat metric $\eta_{\mu \nu}$.

Having determined under what conditions ${\cal{T}}_{\mu \nu}$
vanishes, we now look for implications of (\ref{tmui}). Obviously,
vanishing of ${\cal{T}}_{\mu j}$ and ${\cal{T}}_{i \nu}$ is
guaranteed if  $\phi_0(z)$ in (\ref{phisoln2})  does not involve
mixed terms of $x^{\mu}$ and $y^i$. In other words, the parameters
$a$, $\zeta$ and $p_{\mu}$ must be global constants yet
$b=b(\vec{y})$. The dependence of $b$ on extra dimensions is
rather general; all that is needed is to satisfy equations of
motion self-consistently. For future reference, taking $a>0$ and
$p_{\mu} p^{\mu} - 2 a b(\vec{y})>0$, one notes that
$\widetilde{V}(\phi_0)$ is minimized at $\phi_0 =0$ for $1/4
> \zeta > \zeta_{crit}$, and at $\phi_0=\overline{\phi}$ with
\begin{eqnarray}
\label{mini2} \overline{\phi}=\left( \frac{(D+6) (\zeta -
\zeta_{crit})}{2 \zeta -1} \frac{ 2 {a}}{\eta^{\mu \nu} {p}_{\mu}
{p}_{\nu} - 2 {a} {b}(\vec{y})}\right)^{ \frac{2 \zeta}{1 - 4
\zeta}}
\end{eqnarray}
for $\zeta < \zeta_{crit}$. Clearly, unless the shock wave
propagation in four dimensions is a spherical one, $a\neq 0$, this
very minimum of $\widetilde{V}(\phi_0)$ is neither possible nor
meaningful.

Finally, we analyze implications of a finite ${\cal{T}}_{i j}$. By
construction, ${\cal{T}}_{i j}$ does not vanish and hence extra
space experiences a nontrivial curving. On the field configuration
(\ref{phisoln2}) for which ${\cal T}_{\mu \nu}$, ${\cal{T}}_{i
\nu}$ and ${\cal{T}}_{\mu j}$ vanish identically, equations of
motion for the metric tensor and $\phi_0$ take the form
\begin{eqnarray}
\label{einsteinextra} {\cal{R}}_{i j} &=& \frac{{\cal{T}}_{i j}(\phi_0)}{M_{\star}^{D+2}-\zeta \phi_0^2}\\
\label{newphieqn} g^{i j} \nabla_i \nabla_j  \phi_0 &=& \zeta
{\cal{R}} \phi_0 + V^{\prime}(\phi_0) -
\widetilde{V}^{\prime}(\phi_0) - a D \frac{\zeta}{1-4 \zeta}
\phi_0^{\frac{1-2\zeta}{2 \zeta}}
\end{eqnarray}
where the Ricci tensor is sourced by
\begin{eqnarray}
{\cal{T}}_{i j}(\phi_0) = \partial_{i} \phi_0
\partial_j \phi_0 - \zeta \nabla_i \nabla_j  \phi_0^2 -  \frac{ 4 a \zeta^2}{1-4 \zeta} \phi_0^{\frac{1}{2
\zeta}}\, g_{i j}
\end{eqnarray}
which require $\phi$ to possess the specific solution $\phi_0$
given in (\ref{phisoln2}). A simultaneous solution of
(\ref{einsteinextra}) and (\ref{newphieqn}) completely determines
the curvature scalar:
\begin{eqnarray}
\label{curvature} {\cal{R}}&=& \frac{1}{M_{\star}^{D+2}} \Bigg\{
\left(2- \frac{1}{\zeta}\right) \left(\widetilde{V}(\phi_0) -
V(\phi_0)\right) + \phi_0 \left(\widetilde{V}^{\prime}(\phi_0) -
V^{\prime}(\phi_0)\right)+  a {D} \zeta \phi_0^{\frac{1}{2
\zeta}}\Bigg\}
\end{eqnarray}
which is a measure of the degree to which the extra space is
curved.

Having worked out the question of under what conditions a bulk scalar in $4+D$ dimensions gravitates only in a
subgroup of dimensions, we now turn to a discussion of the role and nature of the self-interaction potential
$V_{{\small \mbox{new}}}(\phi)$ of $\phi(z)$. First of all, $V_{{\small \mbox{new}}}(\phi)$ is the scalar
potential felt by a generic scalar field when the higher dimensional metric obtains the block diagonal structure
in (\ref{gmunu}-\ref{gij}). In other words, it refers to part of the action density when all derivatives with
respect to $x_{\mu}$ are dropped. In fact, it is not more than a rearrangement of the terms involving derivatives
with respect to extra dimensions so that action density looks like a four-dimensional one to facilitate analysis
of ${\cal{T}}_{\mu \nu} = 0$. In particular,  $V_{{\small \mbox{new}}}(\phi)$ has nothing to do with the
effective potential one would obtain by integrating out degrees of freedom associated with extra dimensions. It
is neither a four-dimensional effective potential in the common sense of the word nor a $(4+D)$--dimensional
effective potential; it is a local function of coordinates, and by taking the specific form
$\widetilde{V}(\phi)$, it directly participates in flattening of the four-dimensional spacetime and in curving of
the extra space via the equations of motion (\ref{einsteinextra}) and (\ref{newphieqn}). To stress again,
$\widetilde{V}(\phi)$ is just an analog of (\ref{pot}), and mathematically it is highly useful since its extrema
in (\ref{mini2}) will feature in the next section when we discuss compactification of the extra dimensions.

In summary, the entire dynamical problem has thus reduced to a
self-consistent solution of (\ref{einsteinextra}) and
(\ref{newphieqn}). The unknowns of the problem are the metric
tensor $g_{i j}(\vec{y})$ and $b(\vec{y})$. Once these two
parameters are fixed one obtains a precise description of the
geometry and shape of the extra space. The terms involving
derivatives with respect to $x^{\mu}$ in the original equations of
motion (\ref{einstein}) and (\ref{phieqn}) have been eliminated by
using the explicit expression of $\phi$ in (\ref{phisoln2}). It is
easy to see that, when $b(\vec{y}) = \frac{a}{2} \eta^{i j} y_i
y_j + \eta^{i j} y_i p_j + b_0$, $b_0$ being a constant, all
components of ${\cal{T}}_{i j}$ vanish and entire
$(4+D)$--dimensional spacetime becomes flat, as discussed in
detail in Sec.2.1 above. All other forms of $b(\vec{y})$ lead to a
nontrivial curving of the extra space. In the next section we will
analyze (\ref{einsteinextra}) and (\ref{newphieqn}),  and discuss
their implications for compactification of the extra dimensions.

\section{Spacetime Compactification}
Spontaneous compactification of $(4+D)$--dimensional spacetime
$\textsc{M}^{4+D}$ into a four-dimensional flat spacetime
$\textsc{M}^4$ spanned by the four macroscopic dimensions times a
$D$-dimensional manifold $\textsc{E}^D$ means that  $\textsc{M}^4
\otimes \textsc{E}^D$ is an energetically preferred solution
compared to $\textsc{M}^{4+D}$ \cite{gz,nilles}. The analysis in
Sec.2.2 made it clear that flatness of $\textsc{M}^4$ is governed
by $\widetilde{V}(\phi)$ not by $V(\phi)$. Indeed, $V(\phi)$ is
the self-interaction potential of $\phi$ in $(4+D)$ dimensions
whereas $V_{{\small \mbox{new}}}(\phi)$ is the potential of the
same scalar as seen from a four-dimensional perspective (see
(\ref{yenitmunu}) which has to vanish for flattening the
four-dimensional subspace). In this sense, higher-dimensional
spacetime configuration consisting of a strictly flat
four-dimensional geometry times an extra curved manifold becomes
energetically preferable only at those $\phi_0$ values for which
$\widetilde{V}(\phi_0)$ is a minimum.

As follows from Sec.2.2, by taking $a>0$ and $\eta^{\mu \nu}
p_{\mu} p_{\nu} - 2 a b(\vec{y}) > 0$ for definiteness, the scalar
potential $\widetilde{V}(\phi_0)$ is found to possess two minima:
$\phi_0 = 0$ (for $\zeta > \zeta_{crit}$) and $\phi_0 =
\overline{\phi}$ (for $\zeta < \zeta_{crit}$) given in
(\ref{mini2}). In the minimum of $\widetilde{V}(\phi_0)$ at
$\phi_0=0$, the scalar field equation (\ref{newphieqn}) is
consistently solved if $V(\phi_0) = \widetilde{V}(\phi_0)$ $i.e.$
$V(0) =0$. This, in fact, follows from (\ref{Vnew}) which implies
that $V(\phi_0)$ must be equal to $\widetilde{V}(\phi_0)$ for any
$\vec{y}$ independent $\phi_0$ configuration. With $\phi_0=0$ and
$V(0) =0$, Ricci tensor and curvature scalar are found to vanish
identically, as follows from (\ref{einsteinextra}) and
(\ref{curvature}). It is clear that the whole picture is
consistent since a vanishing $\phi$ possesses a vanishing
${\cal{T}}_{A B}$ if its potential does also vanish at the field
configuration under concern. Consequently, the minimum of
$\widetilde{V}(\phi_0)$ at $\phi_0 = 0$ represents a Ricci-flat
manifold. This, as mentioned at the beginning of Sec. 2, may be
taken to indicate a strictly flat space $i.e.$ $g_{i j} = \eta_{i
j}$. One thus arrives at the conclusion that if
$\widetilde{V}(\phi_0)$ is minimized at $\phi_0 = 0$ and if $V(0)
= 0$ then the resulting spacetime is a $(4+D)$ dimensional
Minkowski spacetime $\textsc{M}^{4+D}$ $i.e.$ there is no
compactification effect at all. The extra space is a strictly flat
manifold as the four-dimensional subspace itself.

For $\zeta < \zeta_{crit}$, the potential $\widetilde{V}(\phi_0)$
is minimized at a nonzero $\phi_0$ value given in (\ref{mini2}).
The dynamical equations governing the compactification process are
(\ref{einsteinextra}) and (\ref{newphieqn}) where now $\phi_0$ is
replaced by $\overline{\phi}$. All one is to do is to solve
dynamical equations for determining $g_{i j}(\vec{y})$ (with
$D(D+1)/2$ independent components) and $b(\vec{y})$ in a
self-consistent fashion. These two must give a complete
description of the shape and topology of the extra space.

\begin{figure}
\vspace*{-0.20in} \hspace*{1.8in}
\begin{minipage}{8in}
\epsfig{file=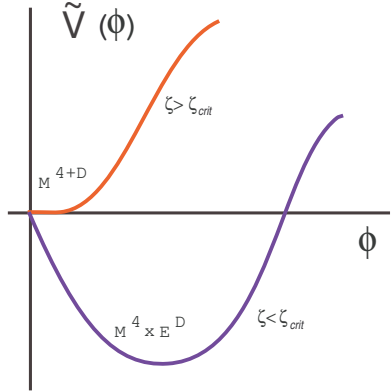,height=2.3in}
\end{minipage}
\vskip .2in \caption{\label{sekil1} {\it The two minima of
$\widetilde{V}(\phi)$ and the corresponding spacetime structures.
}}
\end{figure}

We schematically illustrate the two minima and corresponding
spacetime structures of $\widetilde{V}(\phi)$ in Fig.
\ref{sekil1}. The overall picture is that as $\zeta$ makes a
transition from $\zeta>\zeta_{crit}$ regime to $\zeta <
\zeta_{crit}$ regime the spacetime structure changes from
$\textsc{M}^{4+D}$ to $\textsc{M}^4 \otimes \textsc{E}^D$
spontaneously. The topology and shape of the extra space are
determined by simultaneous solutions of (\ref{einsteinextra}) and
(\ref{newphieqn}) for $\phi_{0} = \overline{\phi}$, defined in
(\ref{mini2}).

An analytic solution of the topology and shape of the extra space
is quite difficult to implement since (\ref{einsteinextra}) and
(\ref{newphieqn}) exhibit a functional dependence on $b(\vec{y})$
and $b(\vec{y})$ itself depends on $g_{i j}(\vec{y})$ via
contraction of the extra coordinates. Therefore, one may
eventually need to resort numerical techniques to determine the
structure of the extra space. Despite these difficulties in
establishing an analytic solution, it may be instructive to
analyze certain simple cases by explicit examples:

{\bf Constant Curvature Space}: The simplest $\overline{\phi}$
configuration which admits an analytic solution of
(\ref{einsteinextra}) and (\ref{newphieqn}) is provided by the
ansatze $b(\vec{y}) = b_0$, a completely $\vec{y}$ independent
configuration. The equation of motion for $\overline{\phi}$
(\ref{newphieqn}) is satisfied with
$V(\overline{\phi})=\widetilde{V}(\overline{\phi})$ as expected
from (\ref{Vnew}). A self-consistent solution of
(\ref{einsteinextra}), (\ref{newphieqn}) and (\ref{curvature})
gives
\begin{eqnarray}
\label{ricci} {\cal{R}}_{i j} = \frac{\cal{R}}{D} g_{i
j}\;\;\;\mbox{with}\;\;\;\; {\cal{R}}=\frac{ a D}{1-4 \zeta}
\overline{\phi}^{\frac{1-4\zeta}{2 \zeta}}
\end{eqnarray}
where vacuum expectation value of the scalar field is fixed via
the consistency condition $M_{\star}^{D+2} = \zeta (1 - 4 \zeta)
\overline{\phi}^2$. In other words, the fundamental scale of
gravity in $(4+D)$ dimensions, $M_{\star}$, fixes the vacuum
expectation value of the bulk scalar $\phi_0$ which is already
designed not to gravitate in the four-dimensional subspace. The
integration constants $a$, $b$ and $p_{\mu}$ in (\ref{phisoln2})
are naturally ${\cal{O}}(M_{\star})$ -- the only mass scale in the
bulk. In fact, by taking $a=\lambda M_{\star}^{2-\frac{1-4
\zeta}{4 \zeta} (D+2)}$ with $\lambda$ being a dimensionless
constant, one finds
\begin{eqnarray}
{\cal{R}}= \lambda D \zeta^{\frac{4 \zeta -1}{4 \zeta}}\ (1-4
\zeta)^{- \frac{1}{4 \zeta}}\ M_{\star}^2
\end{eqnarray}
which is completely determined by $\zeta$, $D$, $\lambda$ and
$M_{\star}$. The resulting spacetime topology is obviously
$\textsc{M}^4 \otimes \textsc{E}^D$ with $\textsc{E}^D$  being a
$D$ dimensional manifold with  positive constant curvature. The
coordinates $\left\{y_i\right\}$ may or may not be compact. The
constant $b(\vec{y})$ case under discussion offers an elegant way
of solving (\ref{einsteinextra}) and (\ref{newphieqn}) and it
results in an intuitively simple interpretation of the manifold
formed by extra dimensions. Indeed, the self-interaction potential
$V(\phi)$, on the partially-gravitating configuration $\phi_0$ in
(\ref{phisoln2}), gets converted into $\widetilde{V}(\phi_0)$
whose minimum at $\phi_0 = \overline{\phi}$ results in a
non-trivial constant-curvature space. In essence, the would-be
cosmological term, $V(\overline{\phi})$, as seen from a
four-dimensional Poincare-invariant perspective via
(\ref{yenitmunu}) is off-loaded and utilized in curving the extra
space (in similarity with the mechanism advocated in \cite{nima2}
for solving the cosmological constant problem).

{\bf More General Cases:} Some further properties of
(\ref{einsteinextra}) and (\ref{newphieqn}) can be revealed by
using an appropriate coordinate system. A suitable setting for
such an analysis is provided by the Riemann normal coordinates
which are defined by a locally-flat space attached to a point $N$
of the manifold of extra dimensions. The local flatness of the
space at (not in any neighborhood of) the point $N$ implies that
$\partial_{i} g_{j k} \equiv 0$ for all $i,j,k = 1,\dots, D$ at
$N$ $i.e.$ all components of the connection coefficients
$\Gamma^{i}_{j k}$ vanish at $N$. Clearly, curvature tensors do
not need to vanish at $N$ since they involve not only
$\Gamma^{i}_{j k}$ but also their first derivatives. Consequently,
one finds
\begin{eqnarray}
{\cal{T}}_{i j}^{(N)}(\overline{\phi}) &=& \frac{4 \zeta^2}{1- 4
\zeta} {\overline{\phi}}^{\frac{1}{2 \zeta}} \left( \frac{1-2
\zeta}{(D+6) (\zeta - \zeta_{crit})}
\partial_i
\partial_j b - a g_{i j}\right)
\end{eqnarray}
so that ${\cal{R}}_{i j}$, unlike (\ref{ricci}) where it is
strictly proportional to $g_{i j}$, now picks up novel structures
generated by $\partial_i \partial_j b$. In other words, it is the
${\vec{y}}$ dependence of $b(\vec{y})$ that enables ${\cal{R}}_{i
j}$ to develop new components not necessarily related to those of
the metric field.

Having replaced covariant derivatives with ordinary ones in this
particular coordinate system, it is now possible to examine
implications of different $\vec{y}$ dependencies of $b(\vec{y})$.
If $b(\vec{y})$ exhibits a linear dependence, $b(\vec{y}) = g^{i
j} p^{\prime}_i y_j$, then the Ricci tensor turns out to depend on
${p^{\prime}}^{k} x^{l} \partial_{i} \partial_{j} g_{k l}$ which
involves curvature tensors rather than the metric tensor itself.
When $b(\vec{y})$ is quadratic in $\vec{y}$, $b(\vec{y})=
(a^{\prime}/2) y_i y^i$, the Ricci tensor now involves $2
a^{\prime} g_{i j} + a^{\prime} x^{k} x^{l} \partial_i
\partial_j g_{k l}$ which again depends on curvature tensors
computed at the point $N$. Consequently, when $b(\vec{y})$
exhibits an explicit $\vec{y}$ dependence the Ricci tensor
involves not only the metric tensor itself (as in (\ref{ricci})
holding for constant-curvature spaces) but also double derivatives
of the metric tensor $i.e.$ the curvature tensors. More general
dependencies are expected to yield more general structures for the
geometry and topology of the extra space.

In general, irrespective of what coordinate system is chosen
$b(\vec{y})$ is a bounded quantity. Therefore, it forces extra
dimensions to take values within a hyperboloid. Indeed, a
quadratic polynomial dependence for $b(\vec{y})$, for instance,
results in
\begin{eqnarray}
\frac{a^{\prime}}{2} y_i y^i + p^{\prime}_i y^i + b_0 < \frac{p_{\mu} p^{\mu}}{ 2 a}
\end{eqnarray}
so that extra dimensions are bounded to have a finite size. For a
purely quadratic dependence one finds $y_i y^i < p_{\mu} p^{\mu} /
a a^{\prime}$ which gives an idea on the maximal size a given
dimension $y^i$ can have. However, for more general, in
particular, non-polynomial $\vec{y}$ dependencies of $b(\vec{y})$
its bounded nature may not imply any size restriction on the extra
space at all. One keeps in mind that all model parameters must
eventually return the correct value of Newton's constant in four
dimensions: $\int d^D y \sqrt{-g} = 8 \pi G_N M_{\star}^{D+2}$.
This constraint requires the extra space to be of finite volume
irrespective of the nature of the manifold
\cite{first,nonabel,gz,gz1,nilles,noncompact}.

\section{Conclusion and Future Prospects}
In this work we have introduced a new method of spontaneous
compactification which involves a partially gravitating bulk
scalar field. We have systematically constructed first a
completely non-gravitating scalar field and then a partially
gravitating one. We have examined scalar field configurations and
minimum energy configurations in each case. Finally, we have
discussed implications of a partially gravitating scalar for
spacetime compactification. Our analysis here serves as an
existence proof of a novel scalar-induced compactification. In
particular, existence of a constant-curvature manifold for extra
dimensions, and other novel properties observed in the frame of
Riemann normal coordinates are particularly encouraging
indications for the fact that a single scalar field, non-minimally
coupled to the curvature scalar, can indeed lead to spontaneous
compactification of the extra dimensions.

Before concluding, it may be useful to discuss briefly some
aspects which have been left untouched in the text. The method of
compactification we have discussed can be straightforwardly
extended to cases with several scalar fields. However, bulk fields
with non-vanishing spin may not always exhibit a physically
sensible configuration when ${\cal T}_{\mu \nu} =0$ is imposed
(for instance, a fermion $\Psi(x^{\mu}, y^i)$ acquires a vanishing
energy-momentum tensor when $\left(\gamma_{\mu} \partial_{\nu} +
\gamma_{\nu}
\partial_{\mu}\right) \Psi = 0$ which comprise the equations of
motion but are much wider than them) . Therefore, we inherently
assume that all fields but $\phi(x^{\mu}, y^i)$ are
long-wavelength modes, and $\phi(x^{\mu}, y^i)$, a gauge singlet
scalar, realizes dynamical compactification at energies
${\cal{O}}(M_{\star})$. The low-energy fields disrupt strict
flatness of $M^{4}$ depending on how their energy scale compares
with $M_{\star}$.

It is necessary to determine a simultaneous solution of
(\ref{einsteinextra}) and (\ref{newphieqn}) for having a precise
knowledge of the shape and topology of the aimed-at manifold. In
particular, these equations cannot be guaranteed to be free of
singularities in the extra space. A detailed analysis is expected
to shed light on nature of such singularities (see, for instance,
\cite{nilles2} for an analysis of the singularities in braneworld
scenarios with a self-tuning cosmological term) . Moreover, a full
account of the spontaneous compactification might require a
numerical determination of variables for sample values of the
parameters. It will be after such an analysis that one will have
detailed information on under what conditions the extra space
takes a given shape and topology.

Another important issue is the determination of excitation
spectrum about the background geometry we have determined. In
other words, it is necessary to determine the gravi-particle
spectra corresponding to normal modes generated by small
oscillations about the background (see \cite{gz1}, for instance).
This involves shifts $\eta_{\mu \nu} \rightarrow \eta_{\mu \nu} +
h_{\mu \nu}$, $g_{i j} \rightarrow g_{i j} + h_{i j}$,
$\phi(x^{\mu}, y^{i}) \rightarrow \phi(x^{\mu}, y^{i}) +
\delta(x^{\mu}, y^{i})$ as well as small but finite values of
$g_{\mu j}$ and $g_{i \nu}$. In doing the spectrum analysis,
particular care should be payed to the fact that the partially
gravitating scalar field configuration in (\ref{phisoln2}) depends
explicitly on the metric tensor, and thus, its variation stems
from both $\delta(x^{\mu}, y^{i})$ and variations of the metric
components.

One final remark concerns the use of higher curvature gravity.
Indeed, higher-curvature gravity theories which generalize
Einstein-Hilbert action to a function $f\left({\cal{R}}, \Box
\cal{R}\right)$ of the curvature scalar can be mapped, via
conformal transformations, into Einstein-Hilbert action plus a
scalar field theory \cite{demir,conf}. In this context, the scalar
field theory which facilitates the compactification may be
interpreted to have pure gravitational origin, and this may entail
possibility of spontaneous compactification via higher curvature
gravity.

These aforementioned points summarize some of the important and
yet-to-be done aspects of the compactification process advocated
in this work. In conclusion, we have shown that a bulk scalar
field in $4+D$ dimensions can lead to a spontaneous
compactification of the extra dimensions without inducing a
classical cosmological constant when it gravitates only in those
dimensions which are to be compactified.

The work of D. A. D. was partially supported by Turkish Academy of
Sciences through GEBIP grant, and by the Scientific and Technical
Research Council of Turkey through project 104T503.

%%%%%%%%%%
%%%%%%%%%%    References
%%%%%%%%%%

\end{document}